\documentclass[twocolumn,showpacs,preprintnumbers,a4,prb]{revtex4}
\usepackage{graphicx}
\usepackage{dcolumn}
\usepackage{amsmath}
\usepackage{amssymb}
\usepackage{bm}

\begin{document}

\title{Ferromagnetic-superconducting hybrid films and their possible applications:
A direct study in a model combinatorial film}

\author{D. Stamopoulos, M. Pissas, and E. Manios}
\affiliation{Institute of Materials Science, NCSR "Demokritos",
153-10, Aghia Paraskevi, Athens, Greece.}
\date{\today}

\begin{abstract}
Model combinatorial films (CFs) which host a pure superconductor
adjacent to a ferromagnetic-superconducting hybrid film (HF) are
manufactured for the study of the influence of ferromagnetic
nanoparticles (FNs) on the nucleation of superconductivity.
Careful resistance measurements were performed {\it
simultaneously} on two different sites of the CFs. Enhancement of
superconductivity and magnetic memory effects were observed only
on the hybrid site of the CFs but were absent on their purely
superconducting part. Our results give direct proof that the FNs
modulate the superconducting order parameter in an efficient and
controlled way giving us the possibility of miscellaneous
practical applications.

\end{abstract}


\maketitle

Ferromagnetic-superconducting hybrid systems where magnetism
coexists with superconductivity have been the subject of intensive
theoretical studies for many
years.\cite{Buzdin84,Buzdin03,Buzdin03B,Aarts97} Advances in
fabrication techniques have recently enabled the reliable
preparation of such systems mainly in the form of
HFs.\cite{Martin97,Morgan98,Terentiev99,Stoll02,Villegas03,Lange03,StamopoulosHF,StamopoulosPRB}
Probably the HF which is most widely studied by recent experiments
is the one consisting of ordered or randomly distributed FNs
embedded in a superconducting
layer.\cite{Martin97,Morgan98,Terentiev99,Stoll02,Villegas03,Lange03,StamopoulosHF,StamopoulosPRB}
The most prominent effect that the FNs impose on the
superconductor is the controlled modulation of the superconducting
order parameter. Under specific conditions superconductivity in a
HF may survive (or can be destroyed) at temperatures or magnetic
fields higher (lower) than that observed in a single
superconductor.\cite{Morgan98,Terentiev99,Stoll02,Lange03,StamopoulosHF,StamopoulosPRB}
In most studies transport measurements were mainly employed to
probe the superconducting fraction of the HF. It was observed that
under certain conditions, depending mainly on the alignment of the
FNs, the resistance in the HF maintained a lower value than the
one observed in a reference pure superconducting layer. Most of
the studies attributed the lowering of the measured resistance to
the enhancement of the {\it bulk} and/or the {\it surface}
critical current that the superconductor may
sustain.\cite{Martin97,Morgan98,Terentiev99,Stoll02,Villegas03,Lange03,StamopoulosHF,StamopoulosPRB}
This specific property makes HFs important for power applications.
Except for current-carrying applications HFs could be also useful
as prototypes for the design of magnetoresistive memory devices or
superconductive spin
valves.\cite{Sangjun97,Tagirov99,Buzdin99,Gu02} In the near future
such devices could serve as data storage elements in a similar way
to other candidate devices which are based on different physical
mechanisms (for example giant magnetoresistance elemental
devices).

\begin{figure}[tbp] \centering%
\includegraphics[angle=0,width=8cm]{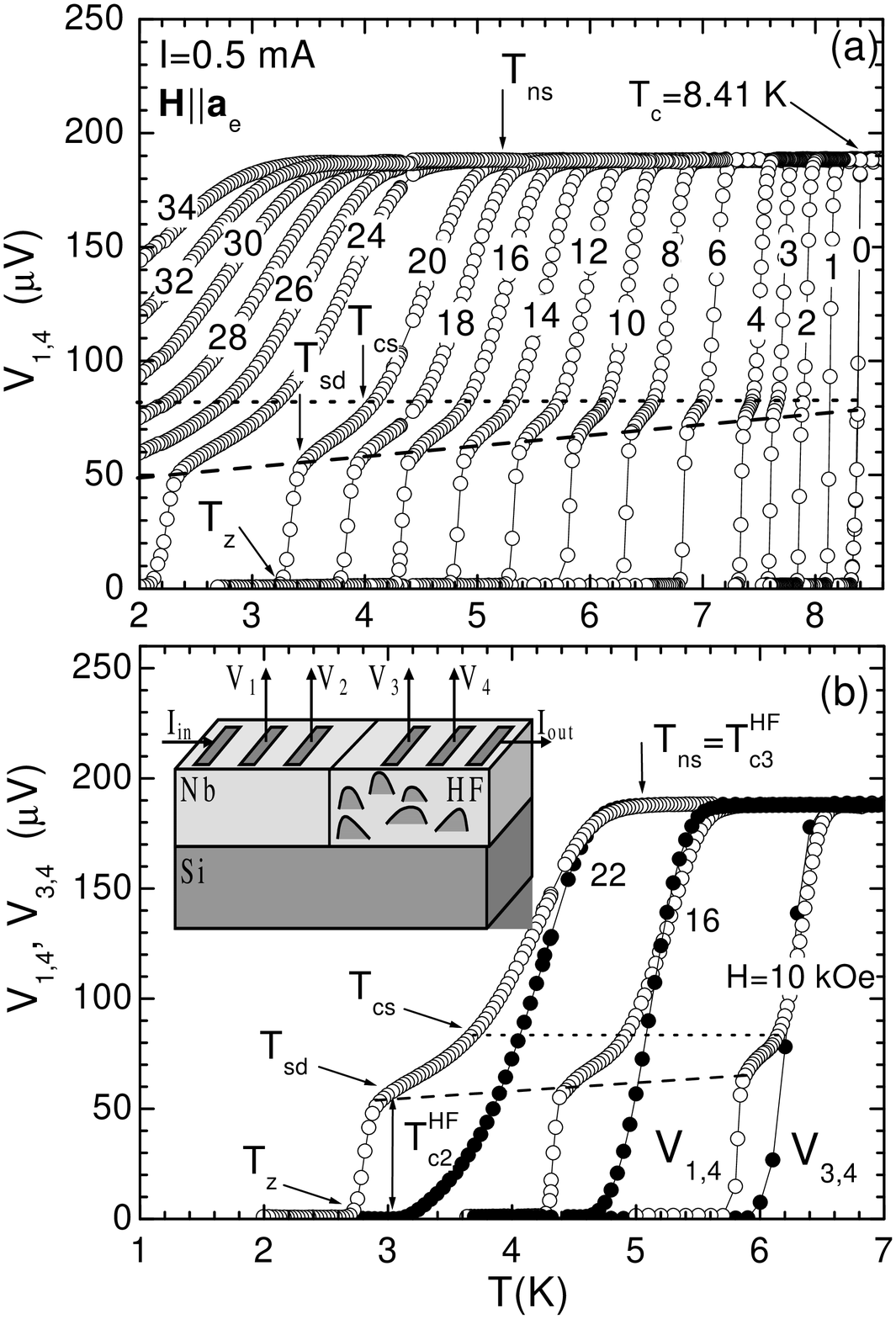}
\includegraphics[angle=0,width=8cm]{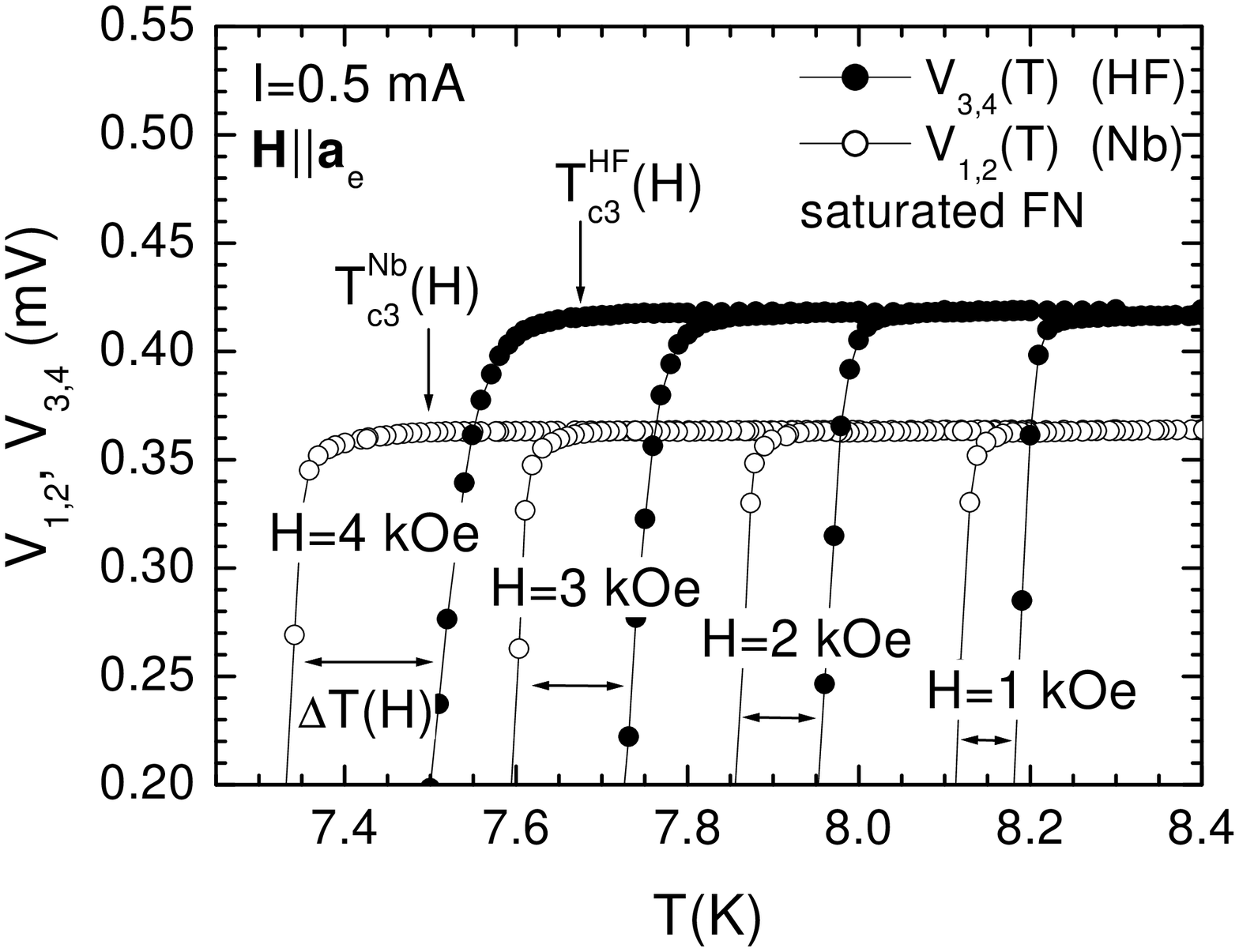}
\caption {(Upper panel) Detailed measurements of the voltage
$V_{\rm 1,4}(T)$ (see the inset) for various magnetic fields $0$
kOe$\leq H\leq 34$ kOe. (Middle panel) Comparative presentation of
voltage curves $V_{\rm 1,4}(T)$ (open circles) and $V_{\rm
3,4}(T)$ (solid circles) at external magnetic fields $H=10,16$ and
$22$ kOe. The inset presents schematically the CF and the
configuration of the performed measurements. (Lower panel)
Comparative presentation of voltage curves $V_{\rm 1,2}(T)$ (open
circles) and $V_{\rm 3,4}(T)$ (solid circles) measured in a second
CF at low magnetic fields $H=1,2,3$ and $4$ kOe. In all cases the
applied field was parallel to the magnetization easy-axis
$\hat{a}_e$ of the FNs.}
\label{b1}%
\end{figure}

In this work we study the nucleation of superconductivity in model
CFs which are constructed by CoPt FNs and a high quality layer of
Nb superconductor. Nb and CoPt FNs were chosen as the ingredients
of the CFs since their respective superconducting and magnetic
properties are well studied and can be modified in a controlled
way by altering the preparation conditions during sputtering and
subsequent
annealing.\cite{Karanasos01,StamopoulosNb,StamopoulosHF,StamopoulosPRB,Manios04}
The FNs employed in this work are anisotropic with their easy-axis
$\hat{a}_e$ of magnetization normal to the surface of the film. In
the constructed CFs the FNs are preferably embedded in only {\it
half} of a high quality Nb layer of thickness $d=200$ nm (see
inset of Fig.\ref{b1}(b), below). In this way the modulation of
the superconducting properties may be studied directly by
performing resistance measurements on the HF and the pure
superconducting areas of the CFs {\it simultaneously} under the
application of the {\it same} dc transport current. The appearance
of spatially modulated superconductivity and site selective memory
effects give direct evidence that in the HF part of the CFs the
nucleation of superconductivity can be greatly modulated by the
dipolar fields of the FNs. The enhancement of the superconducting
regime on the $H-T$ operational diagram suggests that such HFs
could be attractive for current-carrying applications. In
addition, a tristate superconducting magnetoresistive elemental
device could be based on our CF.

Generally, the Nb layers are dc sputtered at a power of $57$ W on
a $2^{''}$ pure Nb target and at an Ar pressure ($99.999$ \% pure)
of $3$ mTorr.\cite{StamopoulosNb} In the present study we managed
to prepare films of even better quality by employing the following
procedure. Since Nb is a strong absorber of oxygen, usually Nb
oxide is grown along grain boundaries resulting in a suppression
of the desired superconducting properties. In order to eliminate
the residual oxygen that possibly existed in the chamber even
after long periods ($2-3$ days) of pumping, we performed
pre-sputtering for very long times. During the pre-sputtering
process all the residual oxygen was absorbed by Nb. As a result we
observed that after a pre-sputtering time of one hour the base
pressure in the chamber was improved almost one order of magnitude
exhibiting a typical value of $8\times 10^{-8}$ Torr. Only when
the desired base pressure had been obtained we started the actual
deposition of the Nb layers.

\begin{figure}[tbp] \centering%
\includegraphics[angle=0,width=8cm]{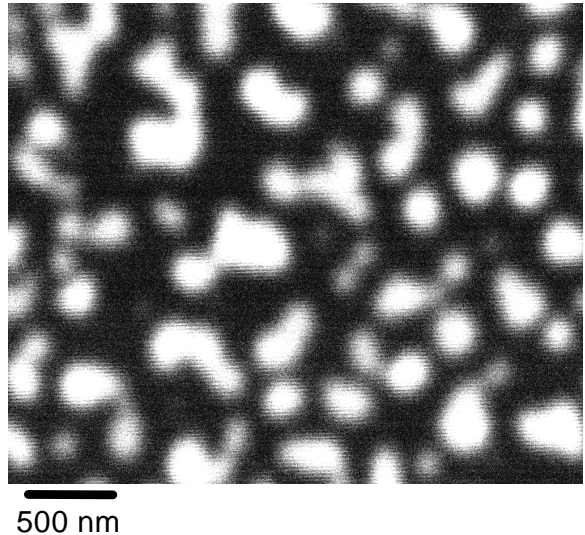}
\caption {Image of the FNs by scanning electron microscopy. The
FNs are isolated and randomly distributed, while their typical
in-plane size and their distance are in the range $100-500$ nm.}
\label{b0}%
\end{figure}

The FNs employed in the present work were prepared as following.
In order to inflict a strong perpendicular anisotropy on the
produced CoPt isolated particles a small quantity of Ag (typical
thickness of the deposited layer $d_{\rm Ag}=0.3-1.2$ nm) should
be used as an extra underlayer.\cite{Karanasos01,Manios04} The Ag
and CoPt layers were deposited on Si$(001)$ substrates by
magnetron sputtering at ambient temperature. The base pressure
before introducing the Ar gas was $5\times 10^{-7}$ Torr and the
pressure during sputtering was $3$ mTorr. The nominal thickness of
Ag layers for the CFs used in this work, was $d_{\rm Ag}=0.6$ nm
and were sputtered by using a dc power of $5$ W on a $2^{''}$ Ag
target at a rate of $1.5$ \AA/sec. The nominal thickness of CoPt
layers was $d_{\rm CoPt}=12$ nm and were sputtered by using an rf
power of $30$ W on a $2^{''}$ CoPt target at a rate of $1.2$
\AA/sec. This specific composition of $d_{\rm Ag}=0.6$ nm and
$d_{\rm CoPt}=12$ nm was chosen for the deposited layers since it
produces the most anisotropic FNs which in addition are isolated
from each other.\cite{Manios04} CoPt films deposited near room
temperature adopt a disordered face centered cubic (fcc) structure
which is magnetically soft. In order to form the hard magnetic and
highly anisotropic $L1o$ phase (ordered face centered tetragonal
or fct) of CoPt, the as deposited films need to be annealed. Thus,
the Ag/CoPt bilayers were annealed for $20$ min at $T=600$ C under
high vacuum ($10^{-7}$ Torr). The annealing process leads to the
formation of self-assembled anisotropic FNs. Their easy-axis
$\hat{a}_e$ of magnetization is normal to the surface of the film.
As already discussed, morphologically, the FNs are isolated and
randomly distributed on the substrate's surface as may be seen in
Fig.\ref{b0}. Typical length scales (in-plane size and distance of
the FNs) are in the range $100-500$ nm. Cross-sectional
transmission electron microscopy data (not shown) revealed that
their thickness is of the order $30-50$ nm. After producing the
FNs the Nb layer was sputtered on top of them according to the
procedure outlined above. The thickness of the deposited Nb layer
is $d\approx 200$ nm. More specific information for the CoPt FNs
may be found in Refs. \onlinecite{Karanasos01,Manios04}.

Our magnetoresistance measurements were performed by applying a dc
transport current (normal to the magnetic field) and measuring the
voltage in the standard four-point configuration. In all
measurements presented below the applied current was $I_{{\rm
dc}}=0.5$ mA, which corresponds to an effective density $J_{{\rm
dc}}\approx 200$ A/cm$^2$ (typical in plane dimensions of the
films are $1\times 4$ (mm)$^2$ to $4\times 4$ (mm)$^2$). The
temperature control and the application of the dc fields were
achieved in a commercial SQUID device (Quantum Design). In all
cases the applied field was parallel to the easy-axis $\hat{a}_e$
of magnetization ($\textbf{H}\parallel \hat{a}_e$ ).

Figures \ref{b1}(a)-\ref{b1}(c) show detailed voltage curves for
two CFs as a function of temperature for various magnetic fields.
The curves presented in Fig.\ref{b1}(a) refer to the voltage
$V_{\rm 1,4}(T)$ which was measured between the characteristic
points $1$ and $4$. As is schematically presented in the inset of
the middle panel these points are positioned on different sites of
the CF. Point $1$ is placed on the superconducting site, while
point $4$ is positioned on the HF part of the CF. Since the FNs
are initially demagnetized, in zero magnetic field the voltage
curve should not exhibit any special feature as indeed is evident
in the data. Once again we note that by employing very long
pre-sputtering times prior to the actual deposition we managed to
produce high-quality Nb layers that maintain homogenous
superconducting properties throughout the CFs. This is evident by
its high zero-field $T_c=8.41$ K and the sharpness of its
single-step transition which exhibits $\Delta T\approx 50$ mK
according to a $10\%-90\%$ criterion (see also Fig.\ref{b3}
below). In contrast to zero-field data, when an external magnetic
field is applied the situation should alter dramatically due to
the presence of the FNs in the HF part. As the external field is
increased any modulation in the critical temperature of the HF, in
respect to the adjacent Nb layer, should result in a structure in
the measured curves $V_{\rm 1,4}(T)$ indicative of two different
transitions. Indeed, this behavior is clearly observed in the data
presented in Fig.\ref{b1}(a). As we perform the measurements in
higher external fields the voltage curves $V_{\rm 1,4}(T)$ broaden
and gradually present a two step feature which indicates the two
different transitions of the two different parts of the CF. More
specifically, we observe that as the temperature decreases the
$V_{\rm 1,4}(T)$ curves deviate from the normal state value at
points $T_{\rm ns}(H)$. In addition they exhibit first a change in
their slope at a field-independent voltage level $V_{\rm
1,4}(T)\simeq 80 \mu V$ (see horizontal dotted line tracing the
points $T_{\rm cs}(H)$) and second a sharp drop towards zero at a
field dependent voltage level (see inclined dashed line tracing
the points $T_{\rm sd}(H)$). Finally, the CF becomes totaly
superconducting at points $T_{\rm z}(H)$ where the resistance gets
zero.

The direct comparison of $V_{\rm 1,4}(T)$ with $V_{\rm 3,4}(T)$
and $V_{\rm 1,2}(T)$ revealed that the upper part of the voltage
curves $V_{\rm 1,4}(T)$ refers to the transition of the HF, while
their lower part refers to the transition of the adjacent Nb
layer. Representative data are shown in Fig.\ref{b1}(b) where we
present comparatively the voltage curves $V_{\rm 1,4}(T)$ and
$V_{\rm 3,4}(T)$ for magnetic fields $H=10, 16$ and $22$ kOe
applied parallel to the easy-axis $\hat{a}_e$ of magnetization of
the FNs. First of all we see that as the temperature decreases
both voltage curves $V_{\rm 1,4}(T)$ and $V_{\rm 3,4}(T)$ deviate
from the normal state value at the same characteristic points
$T_{\rm ns}(H)$.\cite{firstnote} Second, the magnetically
determined bulk upper-critical temperatures $T_{\rm c2}^{\rm
HF}(H)$ of the HF part, represented by the zeroing of the $V_{\rm
3,4}(T)$ curves, \cite{StamopoulosHF,StamopoulosPRB} nicely
correlate with the points where the sharp drop in the voltage
curves $V_{\rm 1,4}(T)$ initiates. We are now in a position to
identify all four characteristic points existing in the $V_{\rm
1,4}(T)$ curves: The points $T_{\rm ns}(H)$ where the normal state
value is obtained refer to the points  $T_{\rm c3}^{\rm HF}(H)$
where surface-like superconductivity starts to form in the HF part
of the CF.\cite{StamopoulosHF,StamopoulosPRB,note} The points
$T_{\rm sd}(H)$ where the sharp drop toward zero initiates (see
inclined dashed line) are related to the bulk upper-critical
temperatures $T_{\rm c2}^{\rm HF}(H)$ of the HF part. The zeroing
temperatures $T_{\rm z}(H)$ reflect the points $T_{\rm c2}^{\rm
Nb}(H)$ where bulk superconductivity is established in the Nb part
of the CF (then superconductivity is maintained in the whole CF)
and finally the points $T_{\rm cs}(H)$ where a change in the slope
is observed in the $V_{\rm 1,4}(T)$ curves (see horizontal dotted
line) refer to the temperatures $T_{\rm c3}^{\rm Nb}(H)$ where the
pure Nb region of the CF enters the normal state as we increase
the temperature.\cite{StamopoulosHF,StamopoulosPRB} {\it An
important outcome of the data presented so far is that in the HF
the superconducting transition takes place at higher temperatures
than in the pure Nb area.} In the lower panel, Fig.\ref{b1}(c) we
focus near the normal state boundary of the $V_{\rm 1,2}(T)$ (open
circles) and $V_{\rm 3,4}(T)$ (solid circles) voltage curves which
were measured at the two different parts of a second CF. These
data show clearly that, for the same applied magnetic field and
current, the normal state boundary of the Nb part is placed at
lower temperatures when compared to the HF part of the CF. Since
all external experimental parameters (temperature, external
magnetic field and applied current) are exactly the same for the
Nb layer existing throughout the CF the enhanced superconducting
temperature of its HF part should be attributed exclusively to the
influence of the FNs.

\begin{figure}[tbp] \centering%
\includegraphics[angle=0,width=7.5cm]{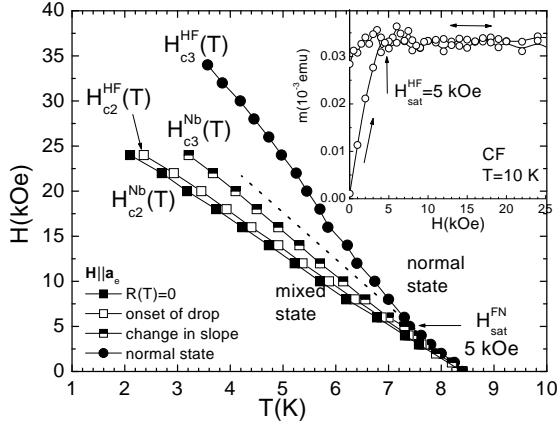}
\caption {The constructed phase diagram for the nucleation of
superconductivity in our CF. The regime where the HF part of the
CF maintains a resistance lower than the normal state value is
greatly enhanced in comparison to the pure Nb part. This is
achieved since an upturn is observed in $H_{\rm c3}^{\rm HF}(T)$
at the saturation field $H_{\rm sat}^{\rm FN}\simeq 5$ kOe of the
FNs (as may be seen in the inset were the initial part of a virgin
loop performed at $T=10$ K is presented). The dotted line
represents the extrapolation of the low-field $H_{\rm c3}^{\rm
HF}(T)$ data in high magnetic fields. The applied magnetic field
was parallel to the easy-axis $\hat{a}_e$ of magnetization
($\textbf{H}\parallel \hat{a}_e$).}
\label{b2}%
\end{figure}

The characteristic points discussed above are summarized in
Fig.\ref{b2} for the field configuration $\textbf{H}\parallel
\hat{a}_e$. In its inset we present the initial part of the virgin
magnetization loop of the CF performed at $T=10$ K. The FNs of the
present film exhibit a saturation field $H_{\rm sat}^{\rm
FN}\simeq 5$ kOe. In the main panel we observe that the line
$H_{\rm c3}^{\rm HF}(T)$, which designates the normal state of the
HF part of the CF, exhibits a change in its slope at the
saturation field $H_{\rm sat}^{\rm FN}\simeq 5$ kOe of the
FNs.\cite{StamopoulosHF,StamopoulosPRB,secondnote} All other
boundary lines referring to the different characteristic points
don't exhibit such a tendency, but clearly maintain an almost
linear behavior in the whole regime investigated in the present
study. As a result the regime where the HF part of the CF
maintains a resistance lower than the normal state value is
significantly enhanced when compared to the respective regime of
the pure Nb part. This finding is in agreement to recent
experimental studies performed in more simple HFs
\cite{StamopoulosHF} and suggests that by controlled variation of
the saturation field of the FNs we may directly enhance the
superconducting regime of the $H-T$ operational
diagram.\cite{StamopoulosPRB} It should also be noted that the
preparation procedure of randomly-distributed isolated FNs is much
simpler compared to the techniques needed for the production of
ordered ones. Thus, randomly distributed FNs could be more
attractive for the construction of HFs that will be used for
commercial current-carrying applications.

The presence of FNs should motivate magnetic memory effects in the
superconducting properties of a HF. To investigate the existence
of such phenomena in our model CF we performed measurements based
on different magnetic histories of the FNs. Such data are
presented in Figs.\ref{b3}(a) and \ref{b3}(b). In Fig. \ref{b3}(a)
solid points refer to the curves $V_{\rm 1,4}(T)$ obtained when
the FNs were initially carefully
demagnetized,\cite{demagnetization} while the open points
(positive fields) and points with crosses (negative fields) refer
to $V_{\rm 1,4}(T)$ data obtained when initially the FNs were
saturated by applying a magnetic field $H>H_{\rm sat}^{\rm
FN}\simeq 5$ kOe. We clearly see that in the temperature regime
$T_{\rm c2}^{\rm HF}(H)<T<T_{\rm c3}^{\rm HF}(H)$ the voltage
curves $V_{\rm 1,4}(T)$ are greatly affected by the magnetic state
of the FNs, while their lower part $T_{\rm c2}^{\rm
Nb}(H)<T<T_{\rm c2}^{\rm HF}(H)$ is left unchanged. This is so
because the lower segment $T_{\rm c2}^{\rm Nb}(H)<T<T_{\rm
c2}^{\rm HF}(H)$ of the curves is related to the transition of the
pure Nb layer of the CF, while the upper segment $T_{\rm c2}^{\rm
HF}(H)<T<T_{\rm c3}^{\rm HF}(H)$ reflects the transition of the HF
part of the CF. {\it Since the superconducting Nb layer extending
in the whole CF is subjected to exactly the same extrinsic
parameters (temperature, external magnetic field and applied
current) it is only the FNs that could motivate the modulation of
superconductivity and the magnetic memory effects observed in its
HF part.} We observe that the curves obtained at positive fields
after the saturation of the FNs are placed in higher temperatures
compared to the curves obtained when the FNs were initially
demagnetized or the ones obtained at negative field values. This
result clearly proves that the nucleation temperature of
superconductivity can be modulated in a controlled way under the
action of FNs.

\begin{figure}[tbp] \centering%
\includegraphics[angle=0,width=7.5cm]{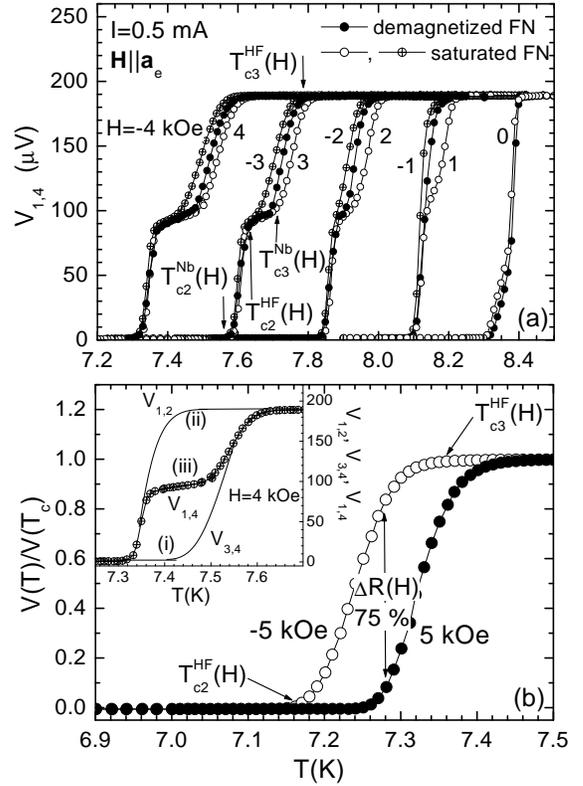}
\caption {(Upper panel) Voltage curves $V_{\rm 1,4}(T)$ at
magnetic fields $H=0,\pm 1,\pm 2,\pm 3$ and $\pm 4$ kOe after
demagnetization (solid points) and after saturation (open points
and points with crosses) of the FNs. Magnetic memory effects are
observed in the $V_{\rm 1,4}(T)$ curves only in the temperature
regime $T_{\rm c2}^{\rm HF}(H)<T<T_{\rm c3}^{\rm HF}(H)$, while
their lower parts $T_{\rm c2}^{\rm Nb}(H)<T<T_{\rm c2}^{\rm
HF}(H)$ coincide entirely. (Lower panel) Normalized voltage
measured in a simple HF at $H=\pm 5$ kOe. The observed change due
to the change in the direction of the external field is $\Delta
V(\pm 5 kOe)=75\%$. The inset presents schematically the three
states of an elemental device which could be based on our CF. In
all cases the magnetic field is parallel to the magnetization
easy-axis $\hat{a}_e$ of the FNs ($\textbf{H}\parallel
\hat{a}_e$).}
\label{b3}%
\end{figure}%

In a recent Letter J.Y. Gu et al. \cite{Gu02} have studied a HF
multilayered device and reported a $25\%$ resistance change which
was motivated by the specific orientation of the magnetization of
the magnetic layers. It should be noted that in their multilayered
HF the observed change in the resistance of the superconducting
layer was not affected by the stray fields of the bordering
magnetic layers but was motivated by the exchange-bias mechanism.
In our HF the observed changes in the voltage are motivated by the
stray fields of the FNs (see below). The data presented in
Fig.\ref{b3}(b) were obtained in a simple HF when the external
field was changed from $H=+5$ kOe to $H=-5$ kOe. A pronounced
percentage change $\Delta V(\pm 5 kOe)=75\%$ is observed in the
measured voltage which is motivated by the magnetic state of the
FNs. Thus, a simple HF could probably serve as a convenient {\it
bistate} magnetoresistive memory device. Furthermore, a CF could
serve as a {\it tristate} memory device which, at constant
temperature, exhibits three distinct resistive states as is
schematically presented in the inset of Fig.\ref{b3}(b): (i) zero
resistance (output at $V_{\rm 3,4}(T)$), (ii) normal state
resistance (output at $V_{\rm 1,2}(T)$) and (iii) an intermediate
value of the resistance (output at $V_{\rm 1,4}(T)$).
Interestingly, the CF can be tuned between states (i), (ii) and
(iii) under the application of an external field and more
importantly its zero-resistance state (i) and intermediate state
(iii) can be additionally modulated by the specific magnetic state
of the FNs. This device could be considered as a development of
the recently proposed magnetoresistive memory and superconducting
spin-valve bistate
elements.\cite{Sangjun97,Tagirov99,Buzdin99,Gu02} It should be
noted that while the operation of those bistate elements
\cite{Sangjun97,Tagirov99,Buzdin99,Gu02} is based on the
exchange-bias mechanism our tristate elemental device simply works
under the action of the dipolar fields of the FNs as discussed
below (see also
Refs.\onlinecite{Lange03,StamopoulosHF,StamopoulosPRB}).

The modulation of superconductivity and the memory effects
observed in our data may be explained by taking into account the
contribution of the internal fields produced by the
FNs.\cite{Lange03,StamopoulosHF,StamopoulosPRB,Cheng99} Briefly,
these internal fields compensate (add to) the external field at
some regimes of the superconductor where the two components are
antiparallel
(parallel).\cite{Cheng99,Lange03,StamopoulosHF,StamopoulosPRB}
Since in these regimes the total effective magnetic field is lower
(higher) than the external field superconductivity will survive
(be destroyed) at temperatures higher (lower) than the ones
observed in a single superconducting
film.\cite{Cheng99,Lange03,StamopoulosHF,StamopoulosPRB} Thus, in
our case when the FNs are initially demagnetized their influence
on the superconducting layer is negligible. As the applied field
increases the FNs are gradually oriented. Their dipolar fields
reduce the external field in the areas of the superconductor which
are placed near their lateral
surfaces.\cite{Lange03,StamopoulosHF,StamopoulosPRB} As a
consequence in these areas of the HF, superconductivity is
preserved at temperatures higher than should be expected in the
absence of FNs. Thus, in our CF the nucleation of
superconductivity should occur at higher temperatures in its HF
part compared to the pure Nb part. This is directly observed in
the results presented in Figs.\ref{b1} and \ref{b3}. Going a step
further, we expect that different behavior should be observed in
the case where the FNs were initially saturated by applying an
external field $H>H_{\rm sat}^{\rm FN}\simeq 5$ kOe and
subsequently the field was decreased to the desired value where
the measurement had to be performed. In this case, during the
measurement all the FNs are in the remanent state throughout the
HF area. As a consequence the suppression of the external magnetic
field by the dipolar fields of the FNs in the areas adjacent to
their lateral surfaces is now more efficient, and more importantly
extends in the whole HF part. As a result, when the FNs are
initially saturated superconductivity should be preserved at
higher temperatures compared to the demagnetized initial state.
This is clearly revealed by the data presented in Fig. \ref{b3}
(a).

Since the discussed effect takes place mainly when all the FNs are
aligned we speculate that at microscopic level the whole process
relies on the formation of percolation paths that trace the
lateral surfaces of the FNs in the whole HF's area. These paths
assist the superconducting component of the transport current and
as a result lower value of the resistance is maintained for higher
temperatures in the HF part of a CF. Unfortunately, above the
saturation field $H_{\rm sat}^{\rm FN}\simeq 5$ kOe all the FN are
saturated so that their ability to compensate the external
magnetic field is entirely expended. As a consequence an
increasing external field will gradually destroy
superconductivity.

In summary, in this article we studied the nucleation of
superconductivity in model CFs consisting of FNs preferably
embedded at the half part of a high quality Nb layer. Resistance
measurements were performed simultaneously at the HF and the
purely superconducting regimes of the CFs under the application of
the same transport current. Under the presence of an external
magnetic field in the HF regime superconductivity is preserved at
temperatures higher than that observed in the pure superconducting
area of the CFs. Magnetic memory effects are selectively observed
only in the HF part but are absent in the pure superconducting
regime. Our results show that in a HF we are able to greatly
enhance the superconducting regime of the $H-T$ operating diagram
thus making a superconductor more attractive for current-carrying
applications. Finally, our CF could be considered as a tristate
elemental magnetoresistive unit and may be useful for the design
of oncoming memory devices.

\pagebreak


\begin{references}
\bibitem{firstnote} In Fig.\ref{b1}(b) it appears that, for
$H=10$ and $16$ kOe, $V_{\rm 3,4}(T)>V_{\rm 1,4}(T)$ near their
onset point, while from its schematic inset it is expected that
$V_{\rm 3,4}(T)<V_{\rm 1,4}(T)$ should always hold. This fact is
not actual, but factitious. During our measurements only the
curves $V_{\rm 1,2}(T)$ and $V_{\rm 3,4}(T)$ could be measured
simultaneously. The measurements of the curves $V_{\rm 1,4}(T)$
were performed in different scans. Thus, the observed irregularity
is related to a small relative temperature shift between $V_{\rm
3,4}(T)$ and $V_{\rm 1,4}(T)$ due to the slightly different values
of the applied magnetic field during different measurements
(no-overshoot operation mode of the SQUID). The data presented in
Fig.\ref{b1}(c), that were obtained when $V_{\rm 1,2}(T)$ and
$V_{\rm 3,4}(T)$ were measured simultaneously, leave no doubt for
the validity of our results.
\bibitem{note} Since the $I-V$ characteristics are non-linear in the regime
$T_{\rm c2}^{\rm Nb}(H)<T<T_{\rm c3}^{\rm HF}(H)$ we attribute
this regime to surface superconductivity  (data not shown here but
may be found in Refs.\onlinecite{StamopoulosHF,StamopoulosPRB}).
\bibitem{secondnote} A second gradual change in the slope of $H_{\rm c3}^{\rm
HF}(T)$ line which is observed around $20$ kOe is not related to
any magnetic characteristic of the FNs. We believe that it is
related to the conventional curvature expected as zero temperature
is approached.
\bibitem{demagnetization} Demagnetization
takes place at $T=10$ K $>T_c=8.41$ K by performing sequential
loops while decreasing the maximum value of the applied magnetic
field. After this procedure is completed the net magnetic moment
of the FN is $m<5\cdot 10^{-6}$ emu.
\end{references}
\end{document}